\documentclass[twocolumn]{aastex62}

\graphicspath{{./}{figures/}}

\received{XXX}
\revised{YYY}
\accepted{ZZZ}
\submitjournal{ApJ}

\usepackage{graphicx}	
\usepackage{amsmath}	
\usepackage{amssymb}	
\usepackage[inline]{enumitem}
\usepackage{gensymb}
\newcommand\lsim{\mathrel{\rlap{\lower4pt\hbox{\hskip1pt$\sim$}}
\raise1pt\hbox{$<$}}}
\newcommand\gsim{\mathrel{\rlap{\lower4pt\hbox{\hskip1pt$\sim$}}
\raise1pt\hbox{$>$}}}

\shorttitle{Eating Planets}
\shortauthors{Stephan et al. 2019}

\begin{document}

\title{Eating Planets for Lunch and Dinner: Signatures of Planet Consumption by Evolving Stars}

\correspondingauthor{Alexander P. Stephan}
\email{alexpstephan@astro.ucla.edu}

\author[0000-0001-8220-0548]{Alexander P. Stephan}
\affil{Department of Physics and Astronomy, University of California, Los Angeles, Los Angeles, CA 90095, USA}
\affiliation{Mani L. Bhaumik Institute for Theoretical Physics, University of California, Los Angeles, Los Angeles, CA 90095, USA}

\author[0000-0002-9802-9279]{Smadar Naoz}
\affiliation{Department of Physics and Astronomy, University of California, Los Angeles, Los Angeles, CA 90095, USA}
\affiliation{Mani L. Bhaumik Institute for Theoretical Physics, University of California, Los Angeles, Los Angeles, CA 90095, USA}

\author[0000-0003-0395-9869]{B. Scott Gaudi}
\affiliation{Department of Astronomy, The Ohio State University, Columbus, OH 43210, USA}

\author{Jesus M. Salas}
\affil{Department of Physics and Astronomy, University of California, Los Angeles, Los Angeles, CA 90095, USA}
\affiliation{Mani L. Bhaumik Institute for Theoretical Physics, University of California, Los Angeles, Los Angeles, CA 90095, USA}



\begin{abstract}

Exoplanets have been observed around stars at all stages of stellar evolution, in many cases orbiting in configurations that will eventually lead to the planets being engulfed or consumed by their host stars, such as Hot Jupiters or ultra-short period planets. Furthermore, objects such as polluted white dwarfs provide strong evidence that the consumption of planets by stars is a common phenomenon. This consumption causes several significant changes in the stellar properties, such as changes to the stellar spin, luminosity, chemical composition, or mass loss processes. Here, we explore this wide variety of effects for a comprehensive range of stellar and planetary masses and stages of stellar evolution, from the main sequence over red giants to the white dwarfs. We determine that planet consumption can cause transient luminosity features that last on the
order of centuries to millennia, and that the post-consumption stellar spins can often reach break-up speeds. Furthermore, stellar moss loss can be caused by this spin-up, as well as through surface grazing interactions, leading to to the formation of unusual planetary nebula shapes or collimated stellar gas ejections. Our results highlight several observable stellar features by which the presence or previous existence of a planet around a given star can be deduced. This will provide future observational campaigns with the tools to better constrain exoplanet demographics, as well as planetary formation and evolution histories.

\end{abstract}


\section{Introduction}\label{sec:intro}
Exoplanets have been observed around a variety of host stars, at all stages of stellar evolution, including main-sequence, subgiant and red giant branch stars \citep[e.g.,][]{Howard+12,Charpinet+11,Barnes+13so,Johnson+11,Gettel+12,Nowak+13,Reffert+15,Niedzielski+15,Niedzielski+16}. Additionally, white dwarf pollution signatures may indicate the presence of planetary systems in a large fraction of white dwarf systems \citep[about $25$ to $50~\%$, e.g.,][]{Jura+2009,Zuckerman+2010,Klein+2010,Klein+2011,Melis+2011}. Planets that stray too close to their host star may get disrupted and finally consumed by the star \citep[e.g., WASP-12b][]{Patra+2017}, leading for example to the observed white dwarf pollution \citep[e.g.,][]{Vanderburg+2015}. 

Dynamical processes play an important role in planetary system formation and evolution, and in particular planet consumption. Interactions between planets may result in orbital instability, possibly plunging planets into the star \citep[e.g.,][]{RasioFord1996,Nag+08,Sourav+08,Naoz11,Tey+13,Denham+2018} Furthermore, the fraction of stellar binaries in the field is high \citep[$\sim 40-70\%$  for $\gsim1$~M$_\odot$ stars, e.g.,][]{Raghavan+10}. The stellar companions may also cause planets to plunge into their host stars \citep[e.g.,][]{LN,Naoz+11sec,Naoz+12bin,Veras+12,Veras+2013,Naoz2016,Veras2016,Veras+2017,Veras+2017_B,Stephan+2017,Stephan+2018,Martinez+2019,Veras+2019}.  
  
The interplay between dynamical effects and post-main-sequence evolution can be very rich. An evolving and expanding star may not only engulf planets on initially close-in orbits, but also far-away planets that have had their eccentricities excited due to perturbations from a third companion. Furthermore, an expanding star will experience stronger tidal forces (which scale with its radius) and may shrink a far-away planet's orbit enough to consume it. In either case, the vicinity to the star will heat the planets significantly prior to contact, turning them into ``Temporary Hot Jupiters'' \citep{Stephan+2018}. However, the mass loss an evolving star undergoes can also expand the orbits, preventing consumption \citep{Valsecchi+2014}. This orbital expansion can change the dynamical stability of a system, especially in the presence of companions, which can lead to star-planet collisions at later times \citep[e.g.,][]{Petrovich+2017,HPZ2016,Stephan+2017}. The interplay of stellar evolution with dynamical processes can therefore explain a variety of interesting observations \citep[e.g.,][]{Xu+2017,Wang+2019,Huber+2019}.

Here we explore the physical processes that a star undergoes as it consumes a planet, for a range of stellar masses and evolutionary phases. The consumption of a planet by a star was considered in the literature as a way to explain a variety of astrophysical phenomena. For example, a planet grazing a stellar surface has been suggested as the cause for the peculiar gas ejections observed from the red giant star V Hydrae \citep{Sahai+2016,Salas+2019}. Furthermore, as a planet enters a star's atmosphere, its interaction with the stellar gas has been suggested to result in transit phenomena such as strong stellar wind, as well as strong optical, UV, and X-ray radiation \citep{Metzger+2012}. Moreover, planet engulfment has also been considered as a cause for non-spherical, dipole-shaped planetary nebulae, similar to the effect in some binary star system \citep[e.g.,][]{Morris1981,MastrodemosMorris1998,Soker1998,Soker+2000,Livio+2002,Morris+2006,Kim+2017}.

When a star consumes a planet, it may have significant effects on the physical properties of the star. For example, it has been shown that the consumption of a Hot Jupiter by a young star can explain some observed patterns of spin-orbit misalignmemts in planetary systems \citep{Matsakos+2015}. Recently, a proof-of-concept calculation for main-sequence G and K-type stars by \citet{Qureshi+18} showed that a consumption of a planet can significantly spin up a star (lowering the spin period). Furthermore, they showed that this spin period change is consistent with the observed bifurcation of spin periods in young open clusters. 

In this work we integrate all the forementioned aspects of planet consumption by stars for a comprehensive range of stellar masses and evolutionary phases. We calculate a range of observational signatures that can be used to infer active planetary consumption events (Section \ref{sec:signatures}). In particular, we determine the phase space of planetary and stellar mass and radius that allow the ejection of stellar gas due to grazing interactions (Section \ref{subsec:grazing}), we calculate the duration and intensity of high-energy UV radiation emitted over the planet's migration through the stellar atmosphere (Section \ref{subsec:lum}), and we estimate the new spin periods of post-consumption stars due to angular momentum conservation (Section \ref{subsec:spin}).

\section{Observable Signatures of Planet Consumption}\label{sec:signatures}

The consumption of planets by stars involves a multitude of processes and effects as a consequence of angular momentum and energy conservation. As a planet begins to graze and contact the stellar surface, gravitational and tidal interactions can disturb or even eject stellar surface material \citep[e.g.,][]{Dosopoulou+17,Salas+2019}. When the planet eventually migrates deeper into the stellar envelope, drag interactions will heat the stellar gas, producing additional luminosity \citep[e.g.,][]{Metzger+2012}, and transfer angular momentum from the planet's orbit onto the star, changing the stellar spin rate and orientation \citep[e.g.,][]{Qureshi+18}. Eventually, the planet will be disrupted and its material will be added to the star, changing its chemical composition. In this section we investigate all of these consumption signatures and determine their strengths and relevance for different stellar types and evolutionary phases.

\subsection{Surface Grazing Interactions}\label{subsec:grazing}

As a planet grazes the surface of a star, it can be expected that stellar surface material will be gravitationally disturbed by the planet, assuming that the planet did not get tidally disrupted beforehand. A planet with radius $R_p$ can approach its host star as close as 
\begin{eqnarray}\label{eq:Roche}
	R_{Roche} &\sim& k R_p \left(\frac{M_*}{M_p}\right)^{1/3} \\ &\sim& k R_\odot \left(\frac{R_p}{R_{Jup}}\right) \left(\frac{M_*}{M_\odot}\right)^{1/3} \left(\frac{M_p}{M_{Jup}}\right)^{-1/3} \nonumber
\end{eqnarray}
without being disrupted, where $M_*$ and $M_p$ are the star's and planet's masses, respectively, and $k$ is a numerical factor on the order of $1.6$ to $2.4$. In general, for any star of solar or heavier mass that has at least slightly evolved towards the later stages or past the main sequence, and which has not yet become a white dwarf or other compact object, this tidal disruption distance is smaller than or on the same order as the radius of the star itself, allowing a planet to reach the stellar surface and to undergo grazing interactions. The star V Hya is very likely an example of this type of interaction.

Over recent decades the carbon star V Hya has been observed to periodically eject ``bullets'' of gaseous material \citep{Sahai+2016}. These ejections can be explained by the close periastron passage of a sub-stellar companion that grazes the stellar surface, scooping up surface material and ejecting some of it as ``bullets'' \citep{Salas+2019}. A variety of mechanisms have been suggested as the cause of the ejections, however here we consider a simple ballistic model \citep[e.g.,][]{Dosopoulou+17}, where the velocity of the bullets $v_b$ is approximately equal to the sum of the planet's periastron passage velocity and the planet's escape velocity
\begin{equation}\label{eq:v_b}
    v_b \sim \sqrt{G\left(M_* + M_p\right)\frac{1+e_p}{a_p\left(1-e_p\right)}} + \sqrt{\frac{2 G M_p}{R_p}} \ ,
\end{equation}
where $a_p$ and $e_p$ are the orbital semi-major axis and eccentricity, respectively.

\begin{figure}
\hspace{0.0\linewidth}
\includegraphics[width=\linewidth]{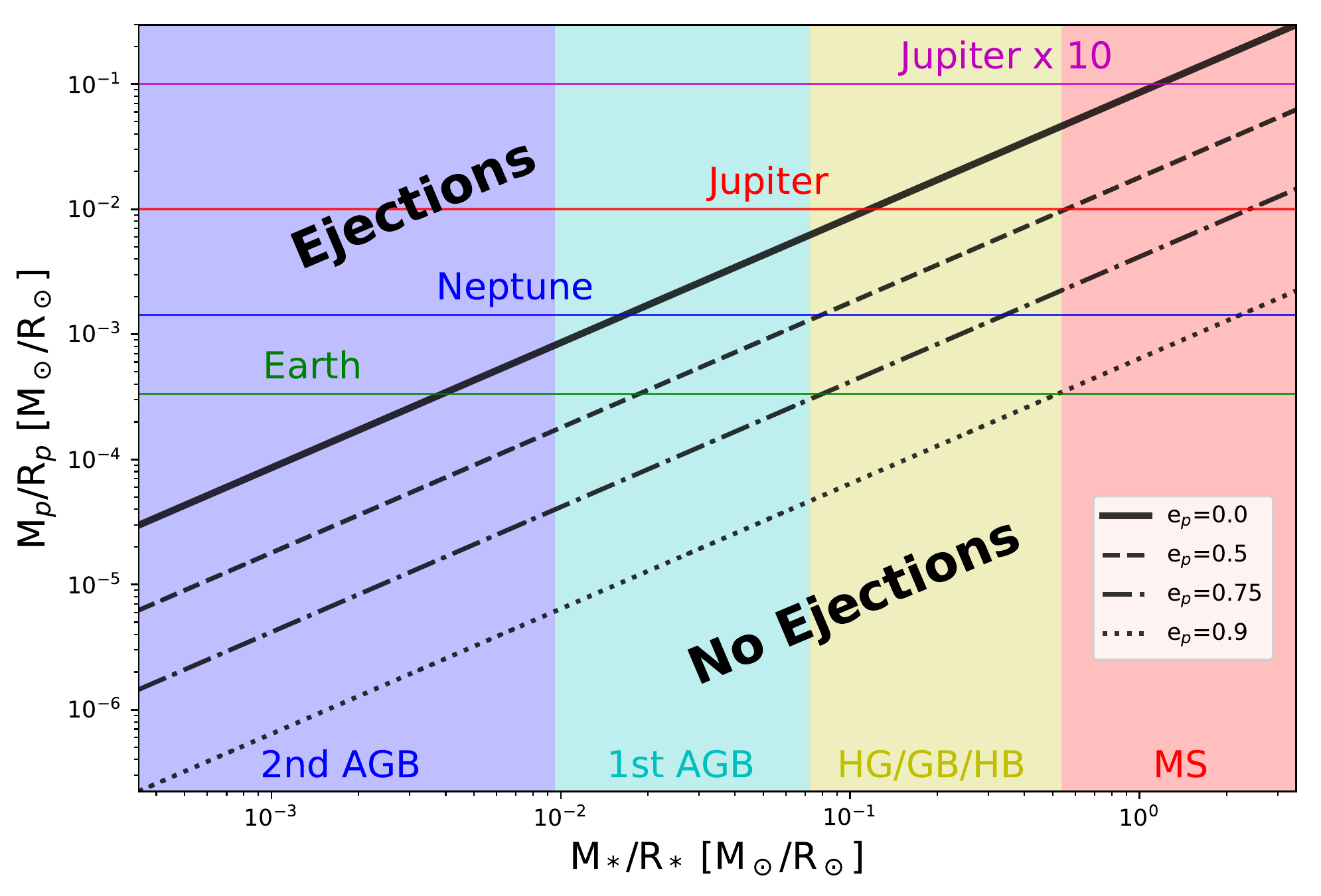}
\caption{{\bf Ejection conditions for a range of planetary masses, stellar types, and orbital eccentricities.} The figure plots the ejection conditions outlined in Equation \ref{eq:ejection}, for four different orbital eccentricities e$_p$: $0$ (solid black line), $0.5$ (dashed black line), $0.75$ (dot-dashed black line), and $0.95$ (dotted black line). Planets whose mass divided by their radius lies above these lines are able to cause V Hya-like ejections for a given star with the corresponding value of its mass divided by its radius. The shaded regions show approximate value ranges of mass divided by radius for different stellar evolutionary phases, for a range of stellar masses. The red region marks values for main-sequence stars, yellow for Hertzsprung gap, giant branch, and helium-burning stars, cyan for 1st Asymptotic Giant Branch (AGB) stars, and blue for 2nd AGB stars. Note that ejections become easier the more evolved a star is, as its ratio of mass to radius decreases. Values for different example planets are shown as vertical lines: Earth in green, Neptune in blue, Jupiter in red, and a massive Jupiter or small Brown Dwarf in magenta. While a Brown dwarf can cause ejections for any type of star or orbital eccentricity, an Earth-sized planet can only cause them for extremely enlarged AGB stars or at extremely high eccentricities. }
\label{fig:ejections}
\end{figure}

However, in order for a bullet to actually leave the star, the bullet velocity must be larger than the stellar escape velocity $v_{esc,*}$ from the periastron, such that 
\begin{equation}\label{eq:escapestar1}
v_b \geq v_{\rm esc,*} \ ,
\end{equation}
where $v_b$ is defined in Equation (\ref{eq:v_b}) and the escape velocity is simply
\begin{equation}
   v_{\rm esc,*}= \sqrt{\frac{2 G M_*}{a_p\left(1-e_p\right)}}.
\end{equation}
Equation (\ref{eq:escapestar1}) can yield a specific relation on the mass to radius ratio of planets and stars. Specifically, assuming that the star is much more massive than the planet and that the periastron distance must be approximately the same as the radius of the star, $R_*$, Equation (\ref{eq:escapestar1}) leads to the condition
\begin{equation}\label{eq:ejection}
    \frac{M_p}{R_p} \geq \frac{M_*}{R_*}\left( \frac{3}{2}+\frac{e_p}{2}-\sqrt{2\left(1+e_p\right)}\right) \ ,
\end{equation}
which needs to be fulfilled for ballistic ejections to efficiently leave the stellar system. Note that for extremely eccentric orbits ($e_p \sim 1$), the right site of Equation \ref{eq:ejection} approaches zero, greatly enhancing ejection likelihood. Figure \ref{fig:ejections} shows how different types of planets are able to cause V Hya-like ejections, depending on their masses and the stellar evolutionary phases. If a planet does not fulfill Equation \ref{eq:ejection}, its interaction with the stellar surface can only eject stellar material onto bound orbits around the star. This could produce a gas and dust disk or cloud around the star whose mass and extend would depend on the planet's mass.

\subsection{Luminosity and Energy Signatures}\label{subsec:lum}

After a planet has grazed a stellar surface, it will eventually orbit fully inside the stellar envelope, shedding orbital energy and angular momentum as it spirals further into the star. The planet will interact with the stellar gas through the drag force $f_d=C_d{\rho_*}{v_k}^2/2$, where $C_d$ is a dimensionless drag coefficient of order unity, ${\rho_*}$ is the stellar density, and $v_k$ is the relative velocity of the planet within the stellar atmosphere \citep{Metzger+2012}. This drag force acts on the planet's effective cross-section $A_p$, which depends on the size of the stellar scale height $H=k_B T/(\mu m_H g)$ (here, $k_B$ is the Boltzmann constant, $T$ is the gas temperature, $\mu$ is the mean molecular mass of the gas, $m_H$ is the mass of a hydrogen atom, and $g$ is the local gravity), as a difference in distance $H$ inwards into the star will increase the density, and thus drag force, by a factor of $e$. If $H$ is much smaller than the planet's radius $R_p$, most of the drag will be caused by the part of the planet most inward into the star, and $A_p$ will be of order ${R_p}^{1/2}H^{3/2}$. The drag torque at a given radius $r$ from the stellar center, ${f_d}{A_p}\times{r}$, will thus contribute to the inward migration of the planet with inward radial speed 
\begin{equation}\label{eq:v_r}
    v_r \sim C_d A_p \frac{ \rho_*(r) }{M_p}  r v_k (r)   \ .
\end{equation}
The inward radial speed due to drag contributes to the rate of orbital energy dissipation via
\begin{equation}\label{eq:Eorbdot}
    \dot{E}_{orb} = \frac{G M_* M_p v_r}{2 r^2} \ ,
\end{equation}
\citep{Metzger+2012}. This energy is added to the stellar gas, heating it up. In general, the travel speed $v_k$ of the planet upon entering the star will be much larger than the stellar atmosphere's sound speed $c_s = \sqrt{\gamma P_*/\rho_*}$; here, $\gamma$ is the adiabatic index, with a value of $5/3$ for ideal gases, and $P_*$ is the gas pressure in the star. The planet will therefore produce a strong shock front as it travels through the stellar atmosphere by which it's orbital energy is dissipated\footnote{Note that there is also a contribution to the inward migration speed and energy dissipation due to tidal forces. This energy is added to the bulk of the star, not the shock front.}. Assuming the standard Rankine-Hugoniot jump conditions and a strong shock, we can estimate the temperature of the gas behind the shock front as
\begin{equation}\label{eq:Tshock_simple}
    T_{shock} = A\frac{\mu m_H}{k_B}v_k^2 \ ,
\end{equation}
where A is a numerical factor of order unity depending on the nature of the gas ($\sim3/16$ for an ideal gas). Note that for fully ionized gas in a star $\mu=0.62$; however, the gas in a red giant is generally not fully ionized pre-shock and has a larger value for $\mu$, which has to be taken into account. 

\begin{figure}
\hspace{0.0\linewidth}
\includegraphics[width=\linewidth]{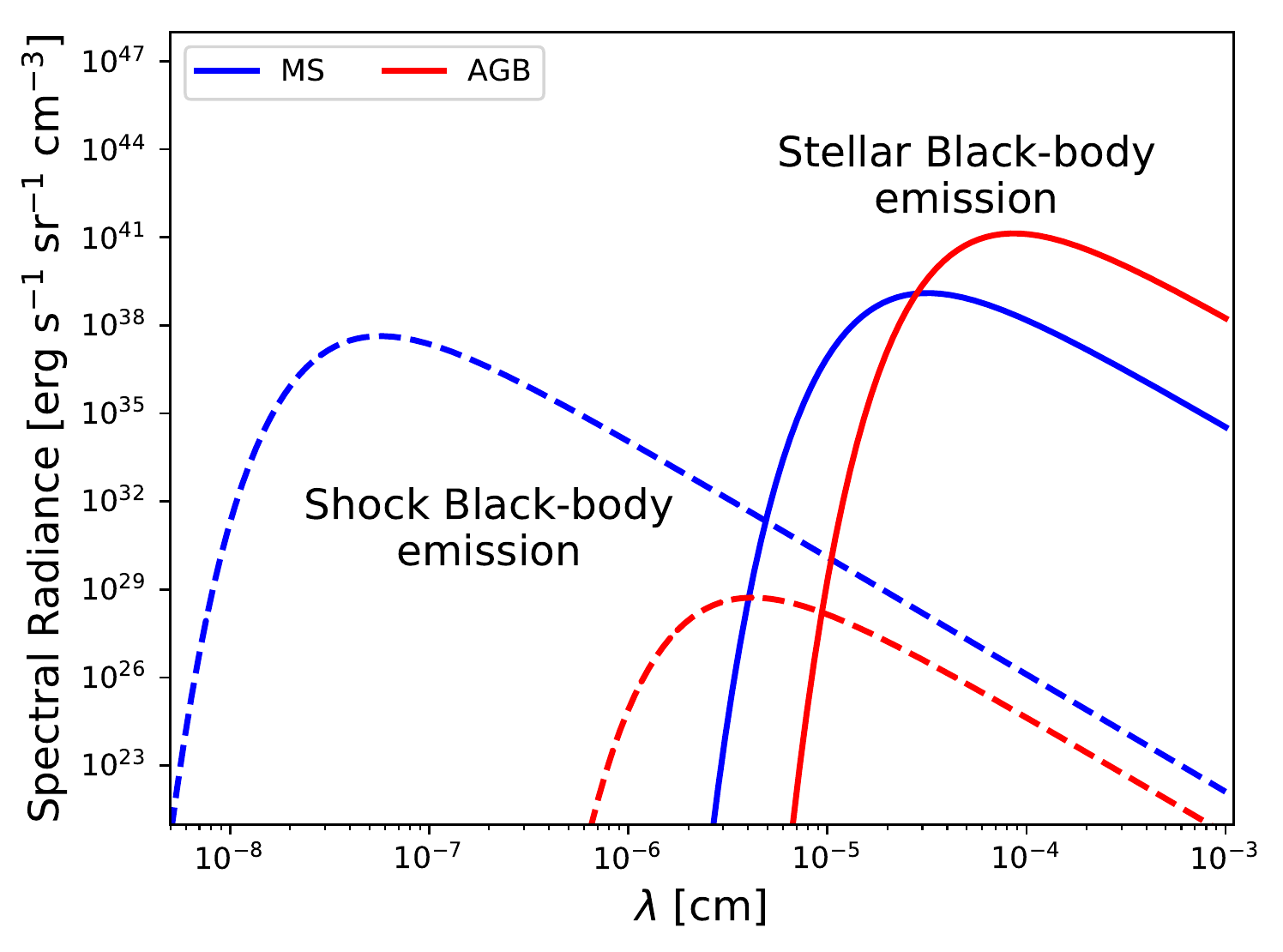}
\caption{{\bf Black-body Emission of a 2 M$_\odot$ Star vs. (un-obscured) Shock Front Black-body Emission during the Main-Sequence and AGB Phases.} The figure shows the black-body emission spectrum approximating the emission of a 2 M$_\odot$ star (solid lines), compared to the black-body emission spectrum approximating the possible emission of the hot shock front as a Jupiter-sized planet is consumed by the star (dashed lines). Shown are the spectra during the main sequence (blue lines) and the AGB phases (red lines). Note that, since the shock is extremely hot, it emits much stronger in the UV and X-ray than the star itself. However, one would probably not actually see most of this emission; during the main sequence the power dissipated in the shock can actually surpass the total luminosity of the star, which should drive a wind that eventually obscures the UV emission, turning it into lower temperature thermal emission \citep{Metzger+2012}. The energy of the shock would also spread and heat surrounding gases as they mix, reducing the emission temperature. During the AGB phase, as the planet migrates further inward into the stellar envelope, the shock's emission is unlikely to penetrate all the way to the surface of the red giant to be observable. As such, these UV signals would most likely be indicators of extremely recent and ongoing consumption events. For example, for a main-sequence star the energy dissipated in the shock should begin to drive a wind and be obscured about 20 years into the merger.}
\label{fig:Radiance}
\end{figure}

The temperature of the shock can be extremely high, potentially reaching several $10,000$ K, producing a lot of X-ray and far-UV radiation, as the peak emission strength will occur at $\lambda_{max}=b_W/T$, where $b_W\sim0.28978$ cm~K is Wien's displacement constant. The radiation intensity as a function of wavelength for blackbodies with the stellar and shock front temperatures are also shown in Figure \ref{fig:Radiance}, following Planck's law. However, the radiation will not necessarily be able to escape the stellar envelope, depending on the optical thickness or ionization state of the gas. Furthermore, the luminosity added to the star can be so large that the immediate region of the shock exceeds its local Eddington limit, which will lead to a wind that will drive an outflow of material from the stellar envelope. This material, in turn, can efficiently block the high energy radiation emitted by the shock \citep{Metzger+2012}, and can also emit additional infrared radiation as it re-radiates intercepted stellar radiation. For different stellar types and phases, the reached temperatures and luminosity emitted by the shock vary, as indicated in Figure \ref{fig:Radiance}. 

\subsection{Angular Momentum Transfer and Stellar Spin-up}\label{subsec:spin}

A planet will transfer angular momentum to its star either through tidal forces, mostly important when the planet is still outside or in the outer regions of the star, or through drag forces, especially important when the planet has reached denser interior layers of the star. The tidal friction timescale of the star can be described by \citep[e.g.,][]{Hut1981,1998KEM,1998EKH,Naoz2016}:
\begin{equation}\label{eq:t_tide}
    t_{TF} = \frac{t_v}{9}\left(\frac{a_p}{R_*}\right)^8 \frac{{M_*}^2}{\left(M_*+M_p\right)M_p} \frac{1}{\left(1+k_l\right)^2} \ ,
\end{equation}
with $t_v$ and $k_l$ being the stellar viscous timescale and Love number, respectively, $a_p$ being the semi-major axis, $R_*$ being the stellar radius, and $M_*$ and $M_p$ being the stellar and planetary masses, respectively. The timescale associated with the gas drag forces outlined in Section \ref{subsec:lum} can be described by 
\begin{equation}\label{eq:t_drag}
    t_{drag} \sim \frac{R_*}{v_r} \sim {C_d}^{-1}\frac{M_p}{\rho_* A_p v_k} \frac{R_*}{a_p} \ ,
\end{equation}
\citep{Metzger+2012}. 

\begin{figure*}
\hspace{0.0\linewidth}
\includegraphics[width=\linewidth]{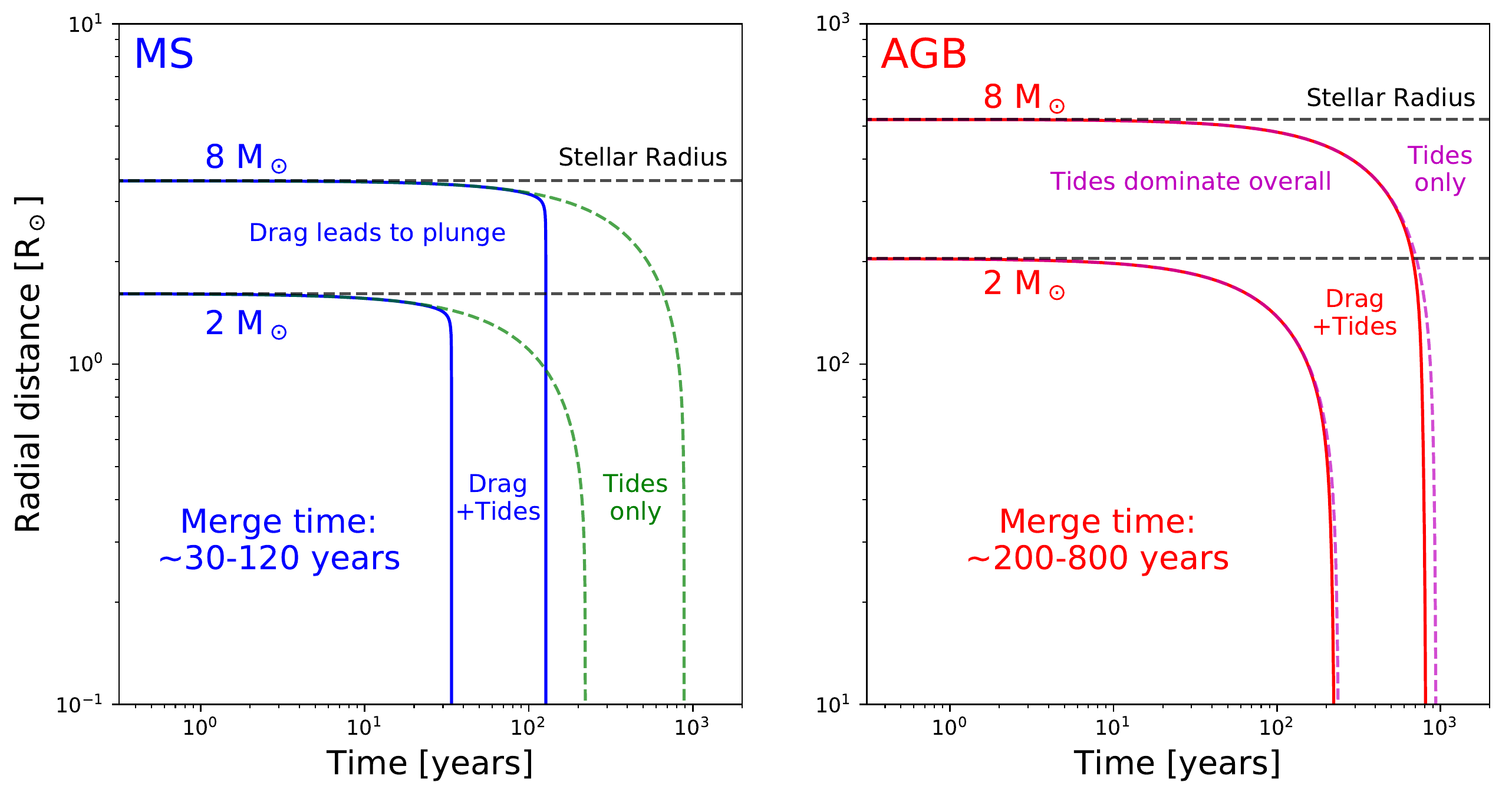}
\caption{{\bf Merging of a Jupiter with a 2 (8) M$_\odot$ Star.} This figure shows the inward radial position over time of a Jupiter-sized planet entering either a main-sequence (MS) star (left panel) or an asymptotic giant branch (AGB) star (right panel). Shown are both the migration due to tides alone (dashed green and magenta lines) and due to drag and tides combined (solid blue and red lines), as well as the radius of the star (dashed black lines). Overall, the planet will have fully merged with the MS star after about 30 to 120 years, and after about 200 to 800 years with the AGB star. The exact time, in all cases, depends strongly on the viscous timescale t$_v$ assumed for the star, as tidal effects dominate while the planets are in the outer layers of the stellar envelope. Here, we assumed t$_v=1.5$ years \citep{Hansen10,Fabrycky+07}. Indeed, while for MS stars drag forces overtake tidal forces after the planets have migrated to about $90~\%$ of the stellar radius, for AGB stars drag forces only begin to dominate at about $50~\%$ of the radius. In either case, once drag forces become dominant, the planets will ``plunge'' into the stellar core on timescales comparable to the planets' orbital periods.}
\label{fig:MSdrag}
\end{figure*}

We can apply the timescales for drag and tidal migration, using realistic models of stellar structure for internal density profiles, to estimate the time needed for a planet to fully merge with a star of given mass and evolutionary stage. Here we assume that the stellar envelope is an $n=3$ polytrope, with an additional compact core in the case of the red giant. While a main-sequence star might only require a few tens of years to fully merge with a planet, a red giant star might need hundreds of years to do the same, as shown in Figure \ref{fig:MSdrag}.

In general, drag forces overcome tidal friction eventually after a planet enters a star, both for main-sequence and red giant stars, as can be seen in Figure \ref{fig:MSdrag}. Indeed, the inward migration speed can become even faster than the orbital speed of the planet, leading to a ``plunge'', with subsequent disruption of the planet and mixing into the stellar core. Overall, however, an engulfed planet spends the vast majority of its time in the outer layers of the star, where tides dominate over drag. This makes the total lifetime of the planet before the ``plunge'' highly dependent on the assumed tidal parameters. Once the planet reaches the ``plunge'' distance, it quickly falls into the core. For a main-sequence star the plunge distance is about a tenth of the stellar radius (from the surface), for a red giant star the plunge distance is at about half the stellar radius, as can be seen in Figure \ref{fig:MSdrag}.

Regardless of the mechanism that contributes most to angular momentum transfer and the planet's migration in the stellar envelope, the angular momentum of the planet's orbit will change the spin of the star. Here, we calculate the new spin rates after such spin-up events and estimate under which conditions the spin-up would actually lead to stellar break-up or envelope loss.

The orbital angular momentum of a planet orbiting a star with semi-major axis $a_p$ and eccentricity $e$ is:
\begin{equation}\label{eq:Lorb}
    J_{orb}=\frac{M_* M_p}{M_*+M_p}\sqrt{G(M_*+M_p)a_p(1-e^2)} \ .
\end{equation}
Assuming that the planet is much less massive than the star, which is reasonable given the mass ratio of Jupiter to the Sun, and that the planet orbit's closest approach distance $a_p(1-e)$ must be the same as the size of the star's Roche limit $R_{*,Roche}$, this equation can be simplified to:
\begin{equation}\label{eq:Lorb2}
    J_{orb}\sim m\sqrt{G M_p R_{*,Roche}(1+e)} \ ,
\end{equation}
where we note that $e$ can only vary between values of $0$ and $1$, thus changing the magnitude of the angular momentum at most by a factor of $\sqrt{2}$. The rotational angular momentum of a spinning star is:
\begin{equation}\label{eq:LS}
    J_*=I_* \Omega_* \ ,
\end{equation}
where I$_*$ is the stellar moment of inertia and $\Omega_*$ its rotation frequency. Here we ignore potential differential rotation profiles. For simplicity we assume that a main-sequence star or a red giant's envelope have a polytropic index of about $3$ and basically reaches all the way from the stellar surface to the stellar core, giving the numerical factor $0.08$ for the angular moment of inertia calculations. From this we determine that the stellar angular momentum is about:
\begin{equation}\label{eq:LSS}
    J_{*,env}=I_{*,env}\Omega_{*,env} \sim 0.08 M_{*,env} R_*^2 \Omega_{*,env} \ ,
\end{equation}
with $I_{*,env}$, $\Omega_{*,env}$, $M_{*,env}$  being the stellar envelope's moment of inertia, rotation rate, and mass, respectively.

\begin{figure*}
\hspace{0.0\linewidth}
\includegraphics[width=\linewidth]{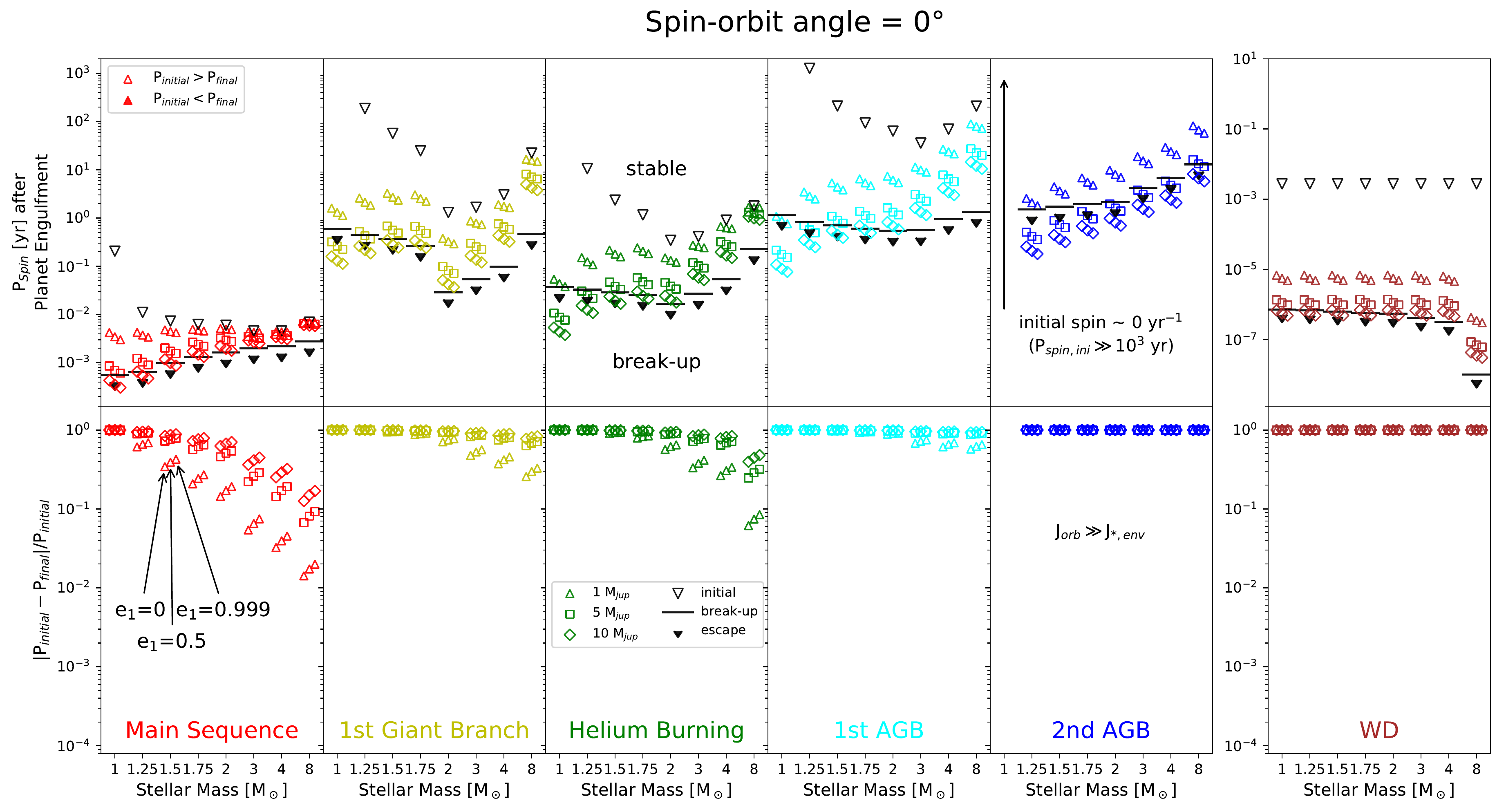}
\caption{{\bf Consumption of prograde-orbiting Planets.} Shown are the spin periods (upper panels) and fractional changes of the spin periods (lower panels) due to the consumption of prograde-orbiting planets of various masses and various eccentricities for a variety of stellar masses and evolutionary phases. The shown evolutionary phases are main-sequence (red), first giant branch (yellow), core helium burning (green), first asymptotic giant branch (cyan), second asymptotic giant branch (blue), and white dwarfs (brown). The empty black downwards triangles show the initial stellar spin periods before consumption, calculated with {\tt SSE} or based on observations in the case of white dwarfs \citep{Kawaler2003}. The tested planetary masses were $1$ M$_{Jup}$ (upwards triangles), $5$ M$_{Jup}$ (squares), and $10$ M$_{Jup}$ (diamonds). For each planetary mass, three orbital eccentricities were tested, shown by groups of three identical symbols; from left to right, the eccentricities were $0$, $0.5$, and $0.999$. The black lines mark the minimum spin periods possible before a star would begin to either lose surface material or be significantly inflated around its equator due to centrifugal forces. The filled-in black downwards triangles show spin periods below which surface material would reach escape speeds, being completely lost from the star. These effects are generally more relevant, as shown, for relatively massive planets being consumed by relatively low-mass stars, and for more evolved stars.}
\label{fig:prograde}
\end{figure*}

\begin{figure*}
\hspace{0.0\linewidth}
\includegraphics[width=\linewidth]{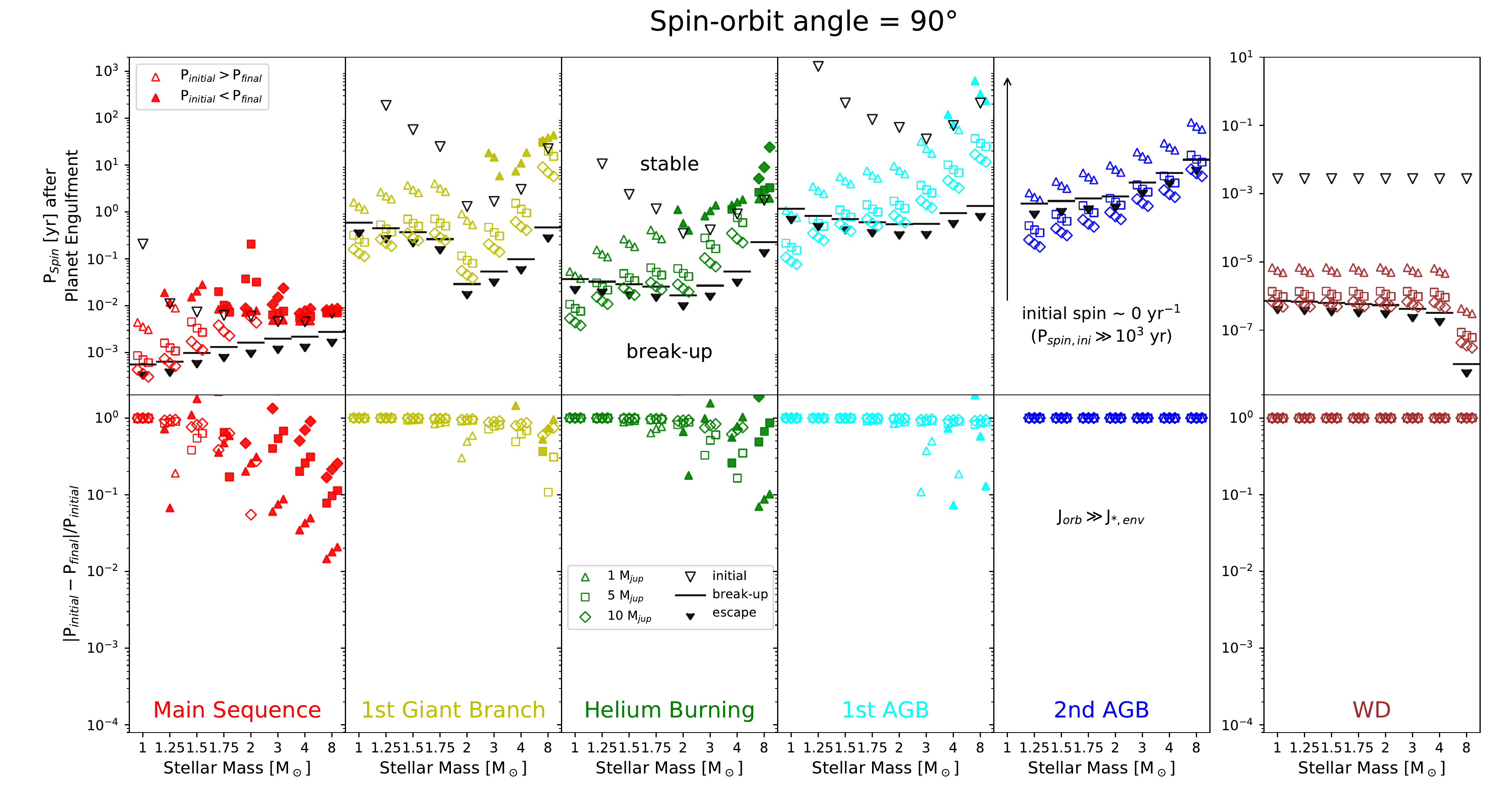}
\caption{{\bf Consumption of retrograde-orbiting Planets.} This figure is similar to Figure \ref{fig:prograde}, however here the planets orbited the stars in retrograde orbits relative to the pre-consumption stellar spin orientation. As a result, it is possible to not just increase a star's spin velocity due to the planets' consumption, but also to instead slow them down. This is shown by the filled-in symbols in the upper and lower panels and appears to be most relevant for massive pre-AGB stars. In all other cases the angular momentum added by the planets overcomes the stars' own angular momentum, resulting in still faster rotation, however in that case the spin direction is reversed relative to the initial spin.}
\label{fig:retrograde}
\end{figure*}

When the star's expanding envelope's Roche limit grows past the planet's orbit, the planet will impart its angular momentum onto the star as it eventually spirals inwards. The angular momenta must add up such that the envelope's new angular momentum is:
\begin{eqnarray}\label{eq:LSSnew}
\mathbf{J}_{*,env,new} &=&  \mathbf{J}_{*,env}+\mathbf{J}_{orb}   \\
&=& 0.08 M_{*,env} {R_*}^2 \mathbf{\Omega}_{*,env,new}  \nonumber \\
&=& {\Omega}_{*,env}+M_p\sqrt{G M_* R_{*,Roche}(1+e)}{\mathbf{\hat{h}}} \ , \nonumber
\end{eqnarray}
assuming here that the stellar radius does not change due to consumption, no differential rotation, and that the planet's mass, as well as the angular momentum of the planetary spin, are negligible. Note also that the stellar Roche limit can be expressed as:
\begin{equation}
    R_{*,Roche}=qR_*\left(\frac{M_*+M_p}{M_*}\right)^{1/3}\sim qR_* \ ,
\end{equation}
with $q$ being a numerical factor assumed here to be about $2.7$. The new spin rate ${\Omega}_{*,env,new}$ is the observable factor, which now becomes:
\begin{equation}\label{eq:OmegaSSnew}
    \mathbf{\Omega}_{*,env,new} = \mathbf{\Omega}_{*,env} + 12.5\times M_p\sqrt{\frac{qGM_*\left(1+e\right)}{M_{*,env}^2 R_*^3}}{\mathbf{\hat{h}}} \ .
\end{equation}
Using Equation \ref{eq:OmegaSSnew}, we can now calculate the changes on a red giant star's spin in a variety of scenarios. We use the stellar evolution code {\tt SSE} \citep{Hurley+00} to evolve stars of masses between $1$ and $8$ M$_\odot$ from the beginning of the main-sequence to their widest possible stellar radius during the Asymptotic Giant Branch (AGB) phase. {\tt SSE} gives information about the stars' radii, masses, core masses, temperatures, and spin rates during all evolutionary phases. We then calculate the changes in spin rates and periods upon consumption of planets with varying masses, orbital eccentricities, and spin-orbit angles. The full ranges of tested parameters are shown in Table \ref{tbl:params}. We compare the changes in spin periods due to consumption during different stellar evolutionary phases, including the Main-Sequence (MS), the first Giant Branch (GB), the Core Helium Burning Phase (HeB), and the 1st and 2nd AGB phases. We note here that generally the stellar radius of a star during the HeB phase is smaller than during the GB phase, however giant planets can still plunge into their host stars during that phase due to effects such as the Eccentric Kozai-Lidov (EKL) mechanism, in which a companion star can induce high orbital eccentricity on a planet due to gravitational perturbations \citep[e.g.,][]{Naoz11,Naoz+12bin,Naoz+12GR,Naoz2016,Stephan+2018}.
\begin{deluxetable}{c r}
\tablecaption{Parameters\label{tbl:params}}
\tabletypesize{\scriptsize}
\tablecolumns{2}
\tablewidth{\columnwidth}
\tablehead{
	\colhead{Parameters} & \colhead{Values}}
\startdata
	{M$_{*}$ [M$_\odot$]} & $1$, $1.25$, $1.5$, $1.75$, $2$, $3$, $4$, $8$ \\
	{M$_{p}$ [M$_{Jupiter}$]} & $1$, $5$, $10$ \\
	{spin-orbit angle [rad]} & $0$, $\pi/2$, $\pi$ \\
	{e$_1$} & $0$, $0.5$, $0.999$
\enddata
\tablecomments{{Listed are the relevant Parameters for Equation \ref{eq:OmegaSSnew} to determine the new stellar spin rate after consumption of a planet.}}
\end{deluxetable}
In Figure \ref{fig:prograde} we show the effects of stellar consumption of prograde-orbiting gas giant planets by stars of different masses and evolutionary phases on the stars' spin periods. In Figure \ref{fig:retrograde} we show the same for a retrograde orbit.

\subsection{Critical Spin Rates and Stellar Break-up}\label{subsec:break}

When calculating the new stellar spins we also need to compare it with the rotational break-up speed of the stars, i.e. the rotation rate at which material on the surface of a given star would be launched into orbit, leaving that surface. The simplest definition of this break-up spin period can be written as:
\begin{equation}\label{eq:break-up}
    P_{*,break-up}=2 \pi \sqrt{\frac{{R_*}^3}{G M_*}}.
\end{equation}
At this spin period a particle on the stellar surface would remain in circular orbit around the star, ignoring potential additional forces such as radiation pressure or similar effects. If the star spins with a smaller period, material from the stellar surface will be launched into orbit. If the spin period is shorter by a factor of $\sqrt{2}$, material launched from the surface will have escape velocity and leave the star completely. At spin periods between $P_{break-up}$ and $P_{break-up}/\sqrt{2}$, material would be launched onto eccentric orbits, effectively leading to the formation of an extreme equatorial bulge and potentially a gaseous disk around the star. As this takes place the spin periods calculated here would obviously need to be adjusted as the shape of the star changes and potentially loses surface material. We note here that calculated spin speeds beyond the break-up speeds are therefore only an indicator for the formation of oblate stars or circumstellar gaseous disks, not for the actual final observable spins.

As Figures \ref{fig:prograde} and \ref{fig:retrograde} show, for a wide variety of stellar and planetary masses the addition of the planetary orbital angular momentum would lead to stellar spins exceeding the spin frequencies required for stellar surface material to be ejected, in particular for more evolved stars past the First Giant Branch, but also some lesser evolved stars. It therefore seems reasonable to assume that such object might already have been observed by previous surveys. Indicators, for example, would be the presence of infrared-excess radiation from debris disks or ejected stellar gas, together with fast stellar spins. Indeed, some observed systems \citep[e.g.,][]{Melis+2009} seem to be good candidates for this process. 

\subsection{Stellar Chemical Enrichment}\label{subsec:chemistry}

The consumption of a planet by a star would also enrich the stellar gas with planetary material. However, at least for gas giants planets, the bulk composition of the planet and the star can be assumed to be very similar, as they are formed from the same protostellar gas and dust disk. Still, there are two main cases where the consumption could produce detectable chemical alterations, lithium enrichment and white dwarf pollution.

Lithium is easily destroyed in stellar nuclear fusion processes through adding a proton to $^7$Li, producing two alpha particles. As such, lithium is heavily depleted in fully convective low-mass stars, where material is continuously mixed back into the core fusion zone, and moderately depleted in more massive stars, where lithium is mostly burned during the pre-main sequence phase but can survive in the stellar atmosphere \citep[e.g.,][]{Thevenin+2017}. However, many observations have shown that a small number of stars, in particular red giants, show abnormally high lithium abundances, with ideas about the cause including dredge-up, new lithium production, or pollution from interstellar gas, brown dwarfs, or planets \citep[e.g.,][]{Alexander1967,Brown+1989,Montalban+2002,Aguilera+2016,BharatKumar+2018,Yan+2018}. Additionally, it has also been suggested that a stellar envelope's metallicity could also be enhanced by absorbing a super-earth \citep{Church+2019}. Naturally, the planetary consumption processes described in this work would enhance stellar lithium or metal abundances as well.

Elements heavier than hydrogen and helium are expected to sink to the cores of white dwarfs, however about a quarter to a third of all white dwarfs still show such heavy elements in their atmospheric spectra \citep[e.g.,][]{Zuckerman+2003,Zuckerman+2010,Koester+2014}. In general it is thought that white dwarfs are being polluted by planetary bodies, usually rocky in composition \citep[e.g.,][]{DebesSigurdsson2002,Jura2003,Jura+2009,Zuckerman+2011,Vanderburg+2015,Xu+2016,Veras+2017_B,Veras+2017}, though some icy bodies containing volatile compounds have also been shown to contribute \citep{Xu+2017,Stephan+2017}. Furthermore, even gas giants could theoretically be brought onto extremely eccentric orbits, getting close enough to the white dwarfs for tidal disruption and eventual pollution \citep{Stephan+2017,Stephan+2018}. The material brought onto a white dwarf from such a massive pollution source could majorly alter the composition of the white dwarf; many white dwarfs are so called helium white dwarfs, with little to no hydrogen left in their atmosphere. A gas giant planet with a mass in the range or $1$ to $10$ M$_{Jup}$, roughly $0.001$ to $0.01$ M$_\odot$, could cover the white dwarf with a hydrogen atmosphere.

\section{Discussion}\label{sec:discussion}

We have studied the variety of consequences due to planetary consumption throughout a star's lifetime. Planetary consumption is expected to be a common outcome of dynamically hot systems \citep[e.g.,][]{RasioFord1996,Nag+08,Sourav+08,Veras,Naoz+12bin,Valsecchi+2014,Petrovich2015,Petrovich+2017,Stephan+2017,Stephan+2018,Denham+2018}. 

Considering a wide range of stellar masses ($1$-$8$~M$_\odot$) and a wide range of planetary masses ($1$~M$_\oplus$-$10$~M$_J$), we examined the effects of planet consumption on a host star. We note here that these calculations are agnostic to the process leading to Roche-limit crossing. The initial stage of star-planet interaction, where the planet interacts with the surface of the star, can have large observable effects. It may lead to ejections of material, either resulting in winds and planetary nebulae \citep[e.g.,][]{Livio+2002}, or even violent, periodic ejections \citep[as was shown for the case of V Hydrae ejections][]{Salas+2019}. The ejection of stellar material due to grazing interactions depends on the star's and planet's mass and size, as well as the orbital eccentricity. We quantified the phase space, considering ballistics ejections\footnote{Note that there are variety of processes that may lead to ejections \citep[e.g.,][]{Goodson+1999,Fendt2003}. Here we adopted the simplest one, which makes no assumptions on magnetic fields or accretion disks.}, at which ejection of material is expected at any given point of a star's lifetime (see Figure \ref{fig:ejections}). In particular, we find that Jupiter mass planets (or higher masses) are efficient in ejecting material over most of the stellar lifetime and for most stellar masses, even for circular orbits. Smaller planets on eccentric orbits can also lead to material ejection rather efficiently. Because the escape velocity from the surface of a star on its 2nd AGB phase is relatively low (a few tens of km/sec), even small, Earth-like planets are sufficient to cause ejections, albeit small ones.

As a planet migrates further into a star's atmosphere it begins to experience gas drag and creates a hot shock as it loses orbital energy. This shock can produce observable UV radiation. For main-sequence stars (blue lines in Figure \ref{fig:Radiance}) the shock's black-body emission intensity can be comparable to the star's radiation \citep[as was shown first in][]{Metzger+2012}. However, the large amount of energy released can drive a wind, obstructing direct observation of the UV emission and converting it into cooler thermal radiation. During the AGB phase the planet's shock radiation is much less intense than the stellar emission. Thus, while it has less power, it may add a small far-UV component to a predominantly visual emission (see red lines in Figure \ref{fig:Radiance}). We estimated the timescales over which the consumption takes place and over which the additional UV signals might be observable. The migration of a planet just inside the surface of a star will mostly be dominated by tides (as shown in Figure \ref{fig:MSdrag}) until drag becomes dominant and ``plunges'' the planet into the stellar core. To reach this phase it takes on the order of a few decades for main-sequence stars and on the order of a few centuries for red giant stars, over which UV signals and shock effects should be visible.

Once the planet is finally consumed by the star we estimated the resulting stellar spin periods from angular momentum conservation\footnote{Note that angular momentum may not be strictly conserved due to mass losses and winds. The consequences of these processes are beyond the scope of this paper.}. We found that main-sequence stars, post consumption, should be rapidly spinning \citep[consistent with][]{Qureshi+18} and that the consumption can significantly alter the star's axis of rotation \citep[consistent with][]{Matsakos+2015}. Red giant stars, post consumption, can often reach spin speeds at or beyond break-up speeds, indicating that these stars would undergo mass-loss from their surface or are strongly tidally distorted into a flattened shape or will become enshrouded by ejected gas. This can also apply to smaller mass stars consuming massive planets or brown dwarfs. Finally, we also showed that white dwarfs can also be significantly spun up, if the angular momentum of a planet can be efficiently transferred to the star. This feature may explain some observed white dwarfs with very short spin periods, such as SDSSJ0837+1856 \citep[e.g.,][]{Hermes+2017,Hermes+2017b}.

Already, observations of short-period planets show that some may have decaying orbits that will eventually let them be consumed by their host star. For example, WASP-12b is a Jupiter size planet that is on a decaying orbit around a $1.35$~M$_\odot$ star \citep[e.g.,][]{Li_nature+2010,Patra+2017}. Based on our calculations (e.g., Figures \ref{fig:prograde} and \ref{fig:retrograde}) we predict that the spin period of WASP-12, upon consumption of this planet, may decrease by a factor of two (prograde orbit), or would be completely flipped (retrograde orbit). Additional planets that are estimated to be on decaying orbits have also been observed \citep[e.g.,][]{Gaudi+2017,MarshallJohnson+2018,LabadieBartz+2019,Johns+2019,Rodriguez+2019}.

While the observation of an active cosumption event might be challenging for a main-sequence star, given the short timescale of a few decades per consumption event, for a red giant star the timescales are relatively favorable, as consumption events would last for centuries or even millenia. Given lifetimes of a few hundred thousend to a few million years for the AGB phase, the chance to observe an engulfment (assuming that every AGB star engulfs a planet) would be on the order of a few tenths to about $1~\%$. Beyond direct observations, the pre- and post-consumption signatures described in this work may provide several avenues to indirectly infer the existence of a planet around a given star. Future HST and JWST observations may detect some of the signatures described here \citep[similar to the gas ejections from V Hydrae described by][]{Sahai+2016}.

\section*{Acknowledgements}

A.P.S. and S.N. acknowledge partial support from the NSF through grant No. AST-1739160. S.N. thanks Howard and Astrid Preston for their generous support.




\bibliographystyle{aasjournal}

\begin{thebibliography}{}
\expandafter\ifx\csname natexlab\endcsname\relax\def\natexlab#1{#1}\fi
\providecommand{\url}[1]{\href{#1}{#1}}

\bibitem[{{Aguilera-G{\'o}mez} {et~al.}(2016){Aguilera-G{\'o}mez},
  {Chanam{\'e}}, {Pinsonneault}, \& {Carlberg}}]{Aguilera+2016}
{Aguilera-G{\'o}mez}, C., {Chanam{\'e}}, J., {Pinsonneault}, M.~H., \&
  {Carlberg}, J.~K. 2016, \apj, 829, 127

\bibitem[{{Alexander}(1967)}]{Alexander1967}
{Alexander}, J.~B. 1967, The Observatory, 87, 238

\bibitem[{{Barnes} {et~al.}(2013){Barnes}, {van Eyken}, {Jackson}, {Ciardi}, \&
  {Fortney}}]{Barnes+13so}
{Barnes}, J.~W., {van Eyken}, J.~C., {Jackson}, B.~K., {Ciardi}, D.~R., \&
  {Fortney}, J.~J. 2013, \apj, 774, 53

\bibitem[{{Bharat Kumar} {et~al.}(2018){Bharat Kumar}, {Singh}, {Eswar Reddy},
  \& {Zhao}}]{BharatKumar+2018}
{Bharat Kumar}, Y., {Singh}, R., {Eswar Reddy}, B., \& {Zhao}, G. 2018, \apjl,
  858, L22

\bibitem[{{Brown} {et~al.}(1989){Brown}, {Sneden}, {Lambert}, \&
  {Dutchover}}]{Brown+1989}
{Brown}, J.~A., {Sneden}, C., {Lambert}, D.~L., \& {Dutchover}, Edward, J.
  1989, \apjs, 71, 293

\bibitem[{{Charpinet} {et~al.}(2011){Charpinet}, {Fontaine}, {Brassard},
  {Green}, {Van Grootel}, {Randall}, {Silvotti}, {Baran}, {{\O}stensen},
  {Kawaler}, \& {Telting}}]{Charpinet+11}
{Charpinet}, S., {Fontaine}, G., {Brassard}, P., {et~al.} 2011, \nat, 480, 496

\bibitem[{{Chatterjee} {et~al.}(2008){Chatterjee}, {Ford}, {Matsumura}, \&
  {Rasio}}]{Sourav+08}
{Chatterjee}, S., {Ford}, E.~B., {Matsumura}, S., \& {Rasio}, F.~A. 2008, \apj,
  686, 580

\bibitem[{{Church} {et~al.}(2019){Church}, {Mustill}, \& {Liu}}]{Church+2019}
{Church}, R.~P., {Mustill}, A.~J., \& {Liu}, F. 2019, arXiv e-prints,
  arXiv:1908.06988

\bibitem[{{Debes} \& {Sigurdsson}(2002)}]{DebesSigurdsson2002}
{Debes}, J.~H., \& {Sigurdsson}, S. 2002, \apj, 572, 556

\bibitem[{{Denham} {et~al.}(2019){Denham}, {Naoz}, {Hoang}, {Stephan}, \&
  {Farr}}]{Denham+2018}
{Denham}, P., {Naoz}, S., {Hoang}, B.-M., {Stephan}, A.~P., \& {Farr}, W.~M.
  2019, \mnras, 482, 4146

\bibitem[{{Dosopoulou} {et~al.}(2017){Dosopoulou}, {Naoz}, \&
  {Kalogera}}]{Dosopoulou+17}
{Dosopoulou}, F., {Naoz}, S., \& {Kalogera}, V. 2017, \apj, 844, 12

\bibitem[{{Eggleton} {et~al.}(1998){Eggleton}, {Kiseleva}, \& {Hut}}]{1998EKH}
{Eggleton}, P.~P., {Kiseleva}, L.~G., \& {Hut}, P. 1998, \apj, 499, 853

\bibitem[{{Fabrycky} {et~al.}(2007){Fabrycky}, {Johnson}, \&
  {Goodman}}]{Fabrycky+07}
{Fabrycky}, D.~C., {Johnson}, E.~T., \& {Goodman}, J. 2007, \apj, 665, 754

\bibitem[{{Fendt}(2003)}]{Fendt2003}
{Fendt}, C. 2003, \aap, 411, 623

\bibitem[{{Gaudi} {et~al.}(2017){Gaudi}, {Stassun}, {Collins}, {Beatty},
  {Zhou}, {Latham}, {Bieryla}, {Eastman}, {Siverd}, {Crepp}, {Gonzales},
  {Stevens}, {Buchhave}, {Pepper}, {Johnson}, {Colon}, {Jensen}, {Rodriguez},
  {Bozza}, {Novati}, {D'Ago}, {Dumont}, {Ellis}, {Gaillard},
  {Jang-Condell}, {Kasper}, {Fukui}, {Gregorio}, {Ito}, {Kielkopf}, {Manner},
  {Matt}, {Narita}, {Oberst}, {Reed}, {Scarpetta}, {Stephens}, {Yeigh},
  {Zambelli}, {Fulton}, {Howard}, {James}, {Penny}, {Bayliss}, {Curtis},
  {Depoy}, {Esquerdo}, {Gould}, {Joner}, {Kuhn}, {Labadie-Bartz}, {Lund},
  {Marshall}, {McLeod}, {Pogge}, {Relles}, {Stockdale}, {Tan}, {Trueblood}, \&
  {Trueblood}}]{Gaudi+2017}
{Gaudi}, B.~S., {Stassun}, K.~G., {Collins}, K.~A., {et~al.} 2017, \nat, 546,
  514

\bibitem[{{Gettel} {et~al.}(2012){Gettel}, {Wolszczan}, {Niedzielski}, {Nowak},
  {Adam{\'o}w}, {Zieli{\'n}ski}, \& {Maciejewski}}]{Gettel+12}
{Gettel}, S., {Wolszczan}, A., {Niedzielski}, A., {et~al.} 2012, \apj, 745, 28

\bibitem[{{Goodson} \& {Winglee}(1999)}]{Goodson+1999}
{Goodson}, A.~P., \& {Winglee}, R.~M. 1999, \apj, 524, 159

\bibitem[{{Hamers} \& {Portegies Zwart}(2016)}]{HPZ2016}
{Hamers}, A.~S., \& {Portegies Zwart}, S.~F. 2016, \mnras, 462, L84

\bibitem[{{Hansen}(2010)}]{Hansen10}
{Hansen}, B.~M.~S. 2010, \apj, 723, 285

\bibitem[{{Hermes} {et~al.}(2017{\natexlab{a}}){Hermes}, {Kawaler}, {Romero},
  {Kepler}, {Tremblay}, {Bell}, {Dunlap}, {Montgomery}, {G{\"a}nsicke},
  {Clemens}, {Dennihy}, \& {Redfield}}]{Hermes+2017}
{Hermes}, J.~J., {Kawaler}, S.~D., {Romero}, A.~D., {et~al.}
  2017{\natexlab{a}}, \apjl, 841, L2

\bibitem[{{Hermes} {et~al.}(2017{\natexlab{b}}){Hermes}, {G{\"a}nsicke},
  {Kawaler}, {Greiss}, {Tremblay}, {Gentile Fusillo}, {Raddi}, {Fanale},
  {Bell}, {Dennihy}, {Fuchs}, {Dunlap}, {Clemens}, {Montgomery}, {Winget},
  {Chote}, {Marsh}, \& {Redfield}}]{Hermes+2017b}
{Hermes}, J.~J., {G{\"a}nsicke}, B.~T., {Kawaler}, S.~D., {et~al.}
  2017{\natexlab{b}}, \apjs, 232, 23

\bibitem[{{Howard} {et~al.}(2012){Howard}, {Marcy}, {Bryson}, {Jenkins},
  {Rowe}, {Batalha}, {Borucki}, {Koch}, {Dunham}, {Gautier}, {Van Cleve},
  {Cochran}, {Latham}, {Lissauer}, {Torres}, {Brown}, {Gilliland}, {Buchhave},
  {Caldwell}, {Christensen-Dalsgaard}, {Ciardi}, {Fressin}, {Haas}, {Howell},
  {Kjeldsen}, {Seager}, {Rogers}, {Sasselov}, {Steffen}, {Basri},
  {Charbonneau}, {Christiansen}, {Clarke}, {Dupree}, {Fabrycky}, {Fischer},
  {Ford}, {Fortney}, {Tarter}, {Girouard}, {Holman}, {Johnson}, {Klaus},
  {Machalek}, {Moorhead}, {Morehead}, {Ragozzine}, {Tenenbaum}, {Twicken},
  {Quinn}, {Isaacson}, {Shporer}, {Lucas}, {Walkowicz}, {Welsh}, {Boss},
  {Devore}, {Gould}, {Smith}, {Morris}, {Prsa}, {Morton}, {Still}, {Thompson},
  {Mullally}, {Endl}, \& {MacQueen}}]{Howard+12}
{Howard}, A.~W., {Marcy}, G.~W., {Bryson}, S.~T., {et~al.} 2012, \apjs, 201, 15

\bibitem[{{Huber} {et~al.}(2019){Huber}, {Chaplin}, {Chontos}, {Kjeldsen},
  {Christensen-Dalsgaard}, {Bedding}, {Ball}, {Brahm}, {Espinoza}, {Henning},
  {Jord{\'a}n}, {Sarkis}, {Knudstrup}, {Albrecht}, {Grundahl}, {Fredslund
  Andersen}, {Pall{\'e}}, {Crossfield}, {Fulton}, {Howard}, {Isaacson},
  {Weiss}, {Handberg}, {Lund}, {Serenelli}, {R{\o}rsted Mosumgaard},
  {Stokholm}, {Bieryla}, {Buchhave}, {Latham}, {Quinn}, {Gaidos}, {Hirano},
  {Ricker}, {Vanderspek}, {Seager}, {Jenkins}, {Winn}, {Antia}, {Appourchaux},
  {Basu}, {Bell}, {Benomar}, {Bonanno}, {Buzasi}, {Campante}, {{\c{C}}elik
  Orhan}, {Corsaro}, {Cunha}, {Davies}, {Deheuvels}, {Grunblatt}, {Hasanzadeh},
  {Di Mauro}, {Garc{\'\i}a}, {Gaulme}, {Girardi}, {Guzik}, {Hon}, {Jiang},
  {Kallinger}, {Kawaler}, {Kuszlewicz}, {Lebreton}, {Li}, {Lucas}, {Lundkvist},
  {Mann}, {Mathis}, {Mathur}, {Mazumdar}, {Metcalfe}, {Miglio}, {Monteiro},
  {Mosser}, {Noll}, {Nsamba}, {Ong}, {{\"O}rtel}, {Pereira}, {Ranadive},
  {R{\'e}gulo}, {Rodrigues}, {Roxburgh}, {Silva Aguirre}, {Smalley},
  {Schofield}, {Sousa}, {Stassun}, {Stello}, {Tayar}, {White}, {Verma},
  {Vrard}, {Y{\i}ld{\i}z}, {Baker}, {Bazot}, {Beichmann}, {Bergmann}, {Bugnet},
  {Cale}, {Carlino}, {Cartwright}, {Christiansen}, {Ciardi}, {Creevey},
  {Dittmann}, {Do Nascimento}, {Van Eylen}, {F{\"u}r{\'e}sz}, {Gagn{\'e}},
  {Gao}, {Gazeas}, {Giddens}, {Hall}, {Hekker}, {Ireland }, {Latouf}, {LeBrun},
  {Levine}, {Matzko}, {Natinsky}, {Page}, {Plavchan}, {Mansouri-Samani},
  {McCauliff}, {Mullally}, {Orenstein}, {Garcia Soto}, {Paegert}, {van Saders},
  {Schnaible}, {Soderblom}, {Szab{\'o}}, {Tanner}, {Tinney}, {Teske}, {Thomas},
  {Trampedach}, {Wright}, {Yuan}, \& {Zohrabi}}]{Huber+2019}
{Huber}, D., {Chaplin}, W.~J., {Chontos}, A., {et~al.} 2019, \aj, 157, 245

\bibitem[{{Hurley} {et~al.}(2000){Hurley}, {Pols}, \& {Tout}}]{Hurley+00}
{Hurley}, J.~R., {Pols}, O.~R., \& {Tout}, C.~A. 2000, \mnras, 315, 543

\bibitem[{{Hut}(1981)}]{Hut1981}
{Hut}, P. 1981, \aap, 99, 126

\bibitem[{{Johns} {et~al.}(2019){Johns}, {Reed}, {Rodriguez}, {Pepper},
  {Stassun}, {Penev}, {Gaudi}, {Labadie-Bartz}, {Fulton}, {Quinn}, {Eastman},
  {Ciardi}, {Hirsch}, {Stevens}, {Stevens}, {Oberst}, {Cohen}, {Jensen},
  {Benni}, {Villanueva}, {Murawski}, {Bieryla}, {Latham}, {Vanaverbeke},
  {Dubois}, {Rau}, {Logie}, {Rauenzahn}, {Wittenmyer}, {Zambelli}, {Bayliss},
  {Beatty}, {Collins}, {Col{\'o}n}, {Curtis}, {Evans}, {Gregorio}, {James},
  {Depoy}, {Johnson}, {Joner}, {Kasper}, {Khakpash}, {Kielkopf}, {Kuhn},
  {Lund}, {Manner}, {Marshall}, {McLeod}, {Penny}, {Relles}, {Siverd},
  {Stephens}, {Stockdale}, {Tan}, {Trueblood}, {Trueblood}, \&
  {Yao}}]{Johns+2019}
{Johns}, D., {Reed}, P.~A., {Rodriguez}, J.~E., {et~al.} 2019, \aj, 158, 78

\bibitem[{{Johnson} {et~al.}(2011){Johnson}, {Apps}, {Gazak}, {Crepp},
  {Crossfield}, {Howard}, {Marcy}, {Morton}, {Chubak}, \&
  {Isaacson}}]{Johnson+11}
{Johnson}, J.~A., {Apps}, K., {Gazak}, J.~Z., {et~al.} 2011, \apj, 730, 79

\bibitem[{{Johnson} {et~al.}(2018){Johnson}, {Rodriguez}, {Zhou}, {Gonzales},
  {Cargile}, {Crepp}, {Penev}, {Stassun}, {Gaudi}, {Col{\'o}n}, {Stevens},
  {Strassmeier}, {Ilyin}, {Collins}, {Kielkopf}, {Oberst}, {Maritch}, {Reed},
  {Gregorio}, {Bozza}, {Calchi Novati}, {D'Ago}, {Scarpetta},
  {Zambelli}, {Latham}, {Bieryla}, {Cochran}, {Endl}, {Tayar}, {Serenelli},
  {Silva Aguirre}, {Clarke}, {Martinez}, {Spencer}, {Trump}, {Joner}, {Bugg},
  {Hintz}, {Stephens}, {Arredondo}, {Benzaid}, {Yazdi}, {McLeod}, {Jensen},
  {Hancock}, {Sorber}, {Kasper}, {Jang-Condell}, {Beatty}, {Carroll},
  {Eastman}, {James}, {Kuhn}, {Labadie-Bartz}, {Lund}, {Mallonn}, {Pepper},
  {Siverd}, {Yao}, {Cohen}, {Curtis}, {DePoy}, {Fulton}, {Penny}, {Relles},
  {Stockdale}, {Tan}, \& {Villanueva}}]{MarshallJohnson+2018}
{Johnson}, M.~C., {Rodriguez}, J.~E., {Zhou}, G., {et~al.} 2018, \aj, 155, 100

\bibitem[{{Jura}(2003)}]{Jura2003}
{Jura}, M. 2003, \apjl, 584, L91

\bibitem[{{Jura} {et~al.}(2009){Jura}, {Muno}, {Farihi}, \&
  {Zuckerman}}]{Jura+2009}
{Jura}, M., {Muno}, M.~P., {Farihi}, J., \& {Zuckerman}, B. 2009, \apj, 699,
  1473

\bibitem[{{Kawaler}(2003)}]{Kawaler2003}
{Kawaler}, S.~D. 2003, arXiv e-prints, astro

\bibitem[{{Kim} {et~al.}(2017){Kim}, {Trejo}, {Liu}, {Sahai}, {Taam}, {Morris},
  {Hirano}, \& {Hsieh}}]{Kim+2017}
{Kim}, H., {Trejo}, A., {Liu}, S.-Y., {et~al.} 2017, Nature Astronomy, 1, 0060

\bibitem[{{Kiseleva} {et~al.}(1998){Kiseleva}, {Eggleton}, \&
  {Mikkola}}]{1998KEM}
{Kiseleva}, L.~G., {Eggleton}, P.~P., \& {Mikkola}, S. 1998, MNRAS, 300, 292

\bibitem[{{Klein} {et~al.}(2011){Klein}, {Jura}, {Koester}, \&
  {Zuckerman}}]{Klein+2011}
{Klein}, B., {Jura}, M., {Koester}, D., \& {Zuckerman}, B. 2011, \apj, 741, 64

\bibitem[{{Klein} {et~al.}(2010){Klein}, {Jura}, {Koester}, {Zuckerman}, \&
  {Melis}}]{Klein+2010}
{Klein}, B., {Jura}, M., {Koester}, D., {Zuckerman}, B., \& {Melis}, C. 2010,
  \apj, 709, 950

\bibitem[{{Koester} {et~al.}(2014){Koester}, {G{\"a}nsicke}, \&
  {Farihi}}]{Koester+2014}
{Koester}, D., {G{\"a}nsicke}, B.~T., \& {Farihi}, J. 2014, \aap, 566, A34

\bibitem[{{Labadie-Bartz} {et~al.}(2019){Labadie-Bartz}, {Rodriguez},
  {Stassun}, {Ciardi}, {Penev}, {Johnson}, {Gaudi}, {Col{\'o}n}, {Bieryla},
  {Latham}, {Pepper}, {Collins}, {Evans}, {Relles}, {Siverd}, {Bento}, {Yao},
  {Stockdale}, {Tan}, {Zhou}, {Eastman}, {Albrow}, {Bayliss}, {Beatty},
  {Berlind}, {Bozza}, {Calkins}, {Cohen}, {Curtis}, {Esquerdo}, {Feliz},
  {Fulton}, {Gregorio}, {James}, {Jensen}, {Johnson}, {Johnson}, {Joner},
  {Kasper}, {Kielkopf}, {Kuhn}, {Lund}, {Malpas}, {Manner}, {McCrady},
  {McLeod}, {Oberst}, {Penny}, {Reed}, {Sliski}, {Stephens}, {Stevens},
  {Villanueva}, {Wittenmyer}, {Wright}, \& {Zambelli}}]{LabadieBartz+2019}
{Labadie-Bartz}, J., {Rodriguez}, J.~E., {Stassun}, K.~G., {et~al.} 2019,
  \apjs, 240, 13

\bibitem[{{Li} {et~al.}(2010){Li}, {Miller}, {Lin}, \&
  {Fortney}}]{Li_nature+2010}
{Li}, S.-L., {Miller}, N., {Lin}, D. N.~C., \& {Fortney}, J.~J. 2010, \nat,
  463, 1054

\bibitem[{{Lithwick} \& {Naoz}(2011)}]{LN}
{Lithwick}, Y., \& {Naoz}, S. 2011, \apj, 742, 94

\bibitem[{{Livio} \& {Soker}(2002)}]{Livio+2002}
{Livio}, M., \& {Soker}, N. 2002, \apjl, 571, L161

\bibitem[{{Martinez} {et~al.}(2019){Martinez}, {Stone}, \&
  {Metzger}}]{Martinez+2019}
{Martinez}, M., {Stone}, N.~C., \& {Metzger}, B.~D. 2019, arXiv e-prints,
  arXiv:1906.08788

\bibitem[{{Mastrodemos} \& {Morris}(1998)}]{MastrodemosMorris1998}
{Mastrodemos}, N., \& {Morris}, M. 1998, \apj, 497, 303

\bibitem[{{Matsakos} \& {K{\"o}nigl}(2015)}]{Matsakos+2015}
{Matsakos}, T., \& {K{\"o}nigl}, A. 2015, \apjl, 809, L20

\bibitem[{{Melis} {et~al.}(2011){Melis}, {Farihi}, {Dufour}, {Zuckerman},
  {Burgasser}, {Bergeron}, {Bochanski}, \& {Simcoe}}]{Melis+2011}
{Melis}, C., {Farihi}, J., {Dufour}, P., {et~al.} 2011, \apj, 732, 90

\bibitem[{{Melis} {et~al.}(2009){Melis}, {Zuckerman}, {Song}, {Rhee}, \&
  {Metchev}}]{Melis+2009}
{Melis}, C., {Zuckerman}, B., {Song}, I., {Rhee}, J.~H., \& {Metchev}, S. 2009,
  \apj, 696, 1964

\bibitem[{{Metzger} {et~al.}(2012){Metzger}, {Giannios}, \&
  {Spiegel}}]{Metzger+2012}
{Metzger}, B.~D., {Giannios}, D., \& {Spiegel}, D.~S. 2012, \mnras, 425, 2778

\bibitem[{{Montalb{\'a}n} \& {Rebolo}(2002)}]{Montalban+2002}
{Montalb{\'a}n}, J., \& {Rebolo}, R. 2002, \aap, 386, 1039

\bibitem[{{Morris}(1981)}]{Morris1981}
{Morris}, M. 1981, \apj, 249, 572

\bibitem[{{Morris} {et~al.}(2006){Morris}, {Sahai}, {Matthews}, {Cheng}, {Lu},
  {Claussen}, \& {S{\'a}nchez-Contreras}}]{Morris+2006}
{Morris}, M., {Sahai}, R., {Matthews}, K., {et~al.} 2006, in IAU Symposium,
  Vol. 234, Planetary Nebulae in our Galaxy and Beyond, ed. M.~J. {Barlow} \&
  R.~H. {M{\'e}ndez}, 469--470

\bibitem[{{Nagasawa} {et~al.}(2008){Nagasawa}, {Ida}, \& {Bessho}}]{Nag+08}
{Nagasawa}, M., {Ida}, S., \& {Bessho}, T. 2008, \apj, 678, 498

\bibitem[{{Naoz}(2016)}]{Naoz2016}
{Naoz}, S. 2016, \araa, 54, 441

\bibitem[{{Naoz} {et~al.}(2011){Naoz}, {Farr}, {Lithwick}, {Rasio}, \&
  {Teyssandier}}]{Naoz11}
{Naoz}, S., {Farr}, W.~M., {Lithwick}, Y., {Rasio}, F.~A., \& {Teyssandier}, J.
  2011, \nat, 473, 187

\bibitem[{{Naoz} {et~al.}(2013{\natexlab{a}}){Naoz}, {Farr}, {Lithwick},
  {Rasio}, \& {Teyssandier}}]{Naoz+11sec}
---. 2013{\natexlab{a}}, \mnras, 431, 2155

\bibitem[{{Naoz} {et~al.}(2012){Naoz}, {Farr}, \& {Rasio}}]{Naoz+12bin}
{Naoz}, S., {Farr}, W.~M., \& {Rasio}, F.~A. 2012, \apjl, 754, L36

\bibitem[{{Naoz} {et~al.}(2013{\natexlab{b}}){Naoz}, {Kocsis}, {Loeb}, \&
  {Yunes}}]{Naoz+12GR}
{Naoz}, S., {Kocsis}, B., {Loeb}, A., \& {Yunes}, N. 2013{\natexlab{b}}, \apj,
  773, 187

\bibitem[{{Niedzielski} {et~al.}(2015){Niedzielski}, {Wolszczan}, {Nowak},
  {Adam{\'o}w}, {Kowalik}, {Maciejewski}, {Deka-Szymankiewicz}, \&
  {Adamczyk}}]{Niedzielski+15}
{Niedzielski}, A., {Wolszczan}, A., {Nowak}, G., {et~al.} 2015, \apj, 803, 1

\bibitem[{{Niedzielski} {et~al.}(2016){Niedzielski}, {Villaver}, {Nowak},
  {Adam{\'o}w}, {Kowalik}, {Wolszczan}, {Deka-Szymankiewicz}, {Adamczyk}, \&
  {Maciejewski}}]{Niedzielski+16}
{Niedzielski}, A., {Villaver}, E., {Nowak}, G., {et~al.} 2016, \aap, 588, A62

\bibitem[{{Nowak} {et~al.}(2013){Nowak}, {Niedzielski}, {Wolszczan},
  {Adam{\'o}w}, \& {Maciejewski}}]{Nowak+13}
{Nowak}, G., {Niedzielski}, A., {Wolszczan}, A., {Adam{\'o}w}, M., \&
  {Maciejewski}, G. 2013, \apj, 770, 53

\bibitem[{{Patra} {et~al.}(2017){Patra}, {Winn}, {Holman}, {Yu}, {Deming}, \&
  {Dai}}]{Patra+2017}
{Patra}, K.~C., {Winn}, J.~N., {Holman}, M.~J., {et~al.} 2017, \aj, 154, 4

\bibitem[{{Petrovich}(2015)}]{Petrovich2015}
{Petrovich}, C. 2015, \apj, 799, 27

\bibitem[{{Petrovich} \& {Mu{\~n}oz}(2017)}]{Petrovich+2017}
{Petrovich}, C., \& {Mu{\~n}oz}, D.~J. 2017, \apj, 834, 116

\bibitem[{{Qureshi} {et~al.}(2018){Qureshi}, {Naoz}, \&
  {Shkolnik}}]{Qureshi+18}
{Qureshi}, A., {Naoz}, S., \& {Shkolnik}, E.~L. 2018, \apj, 864, 65

\bibitem[{{Raghavan} {et~al.}(2010){Raghavan}, {McAlister}, {Henry}, {Latham},
  {Marcy}, {Mason}, {Gies}, {White}, \& {ten Brummelaar}}]{Raghavan+10}
{Raghavan}, D., {McAlister}, H.~A., {Henry}, T.~J., {et~al.} 2010, \apjs, 190,
  1

\bibitem[{{Rasio} \& {Ford}(1996)}]{RasioFord1996}
{Rasio}, F.~A., \& {Ford}, E.~B. 1996, Science, 274, 954

\bibitem[{{Reffert} {et~al.}(2015){Reffert}, {Bergmann}, {Quirrenbach},
  {Trifonov}, \& {K{\"u}nstler}}]{Reffert+15}
{Reffert}, S., {Bergmann}, C., {Quirrenbach}, A., {Trifonov}, T., \&
  {K{\"u}nstler}, A. 2015, \aap, 574, A116

\bibitem[{{Rodriguez} {et~al.}(2019){Rodriguez}, {Quinn}, {Huang},
  {Vanderburg}, {Penev}, {Brahm}, {Jord{\'a}n}, {Ikwut-Ukwa}, {Tsirulik},
  {Latham}, {Stassun}, {Shporer}, {Ziegler}, {Matthews}, {Eastman}, {Gaudi},
  {Collins}, {Guerrero}, {Relles}, {Barclay}, {Batalha}, {Berlind}, {Bieryla},
  {Bouma}, {Boyd}, {Burt}, {Calkins}, {Christiansen}, {Ciardi}, {Col{\'o}n},
  {Conti}, {Crossfield}, {Daylan}, {Dittmann}, {Dragomir}, {Dynes}, {Espinoza},
  {Esquerdo}, {Essack}, {Garcia Soto}, {Glidden}, {G{\"u}nther}, {Henning},
  {Jenkins}, {Kielkopf}, {Krishnamurthy}, {Law}, {Levine}, {Lewin}, {Mann},
  {Morgan}, {Morris}, {Oelkers}, {Paegert}, {Pepper}, {Quintana}, {Ricker},
  {Rowden}, {Seager}, {Sarkis}, {Schlieder}, {Sha}, {Tokovinin}, {Torres},
  {Vanderspek}, {Villanueva}, {Villase{\~n}or}, {Winn}, {Wohler}, {Wong},
  {Yahalomi}, {Yu}, {Zhan}, \& {Zhou}}]{Rodriguez+2019}
{Rodriguez}, J.~E., {Quinn}, S.~N., {Huang}, C.~X., {et~al.} 2019, \aj, 157,
  191

\bibitem[{{Sahai} {et~al.}(2016){Sahai}, {Scibelli}, \& {Morris}}]{Sahai+2016}
{Sahai}, R., {Scibelli}, S., \& {Morris}, M.~R. 2016, \apj, 827, 92

\bibitem[{{Salas} {et~al.}(2019){Salas}, {Naoz}, {Morris}, \&
  {Stephan}}]{Salas+2019}
{Salas}, J.~M., {Naoz}, S., {Morris}, M.~R., \& {Stephan}, A.~P. 2019, \mnras,
  487, 3029

\bibitem[{{Soker}(1998)}]{Soker1998}
{Soker}, N. 1998, \aj, 116, 1308

\bibitem[{{Soker} \& {Harpaz}(2000)}]{Soker+2000}
{Soker}, N., \& {Harpaz}, A. 2000, \mnras, 317, 861

\bibitem[{{Stephan} {et~al.}(2018){Stephan}, {Naoz}, \& {Gaudi}}]{Stephan+2018}
{Stephan}, A.~P., {Naoz}, S., \& {Gaudi}, B.~S. 2018, \aj, 156, 128

\bibitem[{{Stephan} {et~al.}(2017){Stephan}, {Naoz}, \&
  {Zuckerman}}]{Stephan+2017}
{Stephan}, A.~P., {Naoz}, S., \& {Zuckerman}, B. 2017, \apjl, 844, L16

\bibitem[{{Teyssandier} {et~al.}(2013){Teyssandier}, {Naoz}, {Lizarraga}, \&
  {Rasio}}]{Tey+13}
{Teyssandier}, J., {Naoz}, S., {Lizarraga}, I., \& {Rasio}, F.~A. 2013, \apj,
  779, 166

\bibitem[{{Th{\'e}venin} {et~al.}(2017){Th{\'e}venin}, {Oreshina}, {Baturin},
  {Gorshkov}, {Morel}, \& {Provost}}]{Thevenin+2017}
{Th{\'e}venin}, F., {Oreshina}, A.~V., {Baturin}, V.~A., {et~al.} 2017, \aap,
  598, A64

\bibitem[{{Valsecchi} {et~al.}(2014){Valsecchi}, {Rasio}, \&
  {Steffen}}]{Valsecchi+2014}
{Valsecchi}, F., {Rasio}, F.~A., \& {Steffen}, J.~H. 2014, \apjl, 793, L3

\bibitem[{{Vanderburg} {et~al.}(2015){Vanderburg}, {Johnson}, {Rappaport},
  {Bieryla}, {Irwin}, {Lewis}, {Kipping}, {Brown}, {Dufour}, {Ciardi}, {Angus},
  {Schaefer}, {Latham}, {Charbonneau}, {Beichman}, {Eastman}, {McCrady},
  {Wittenmyer}, \& {Wright}}]{Vanderburg+2015}
{Vanderburg}, A., {Johnson}, J.~A., {Rappaport}, S., {et~al.} 2015, \nat, 526,
  546

\bibitem[{{Veras}(2016)}]{Veras2016}
{Veras}, D. 2016, Royal Society Open Science, 3, 150571

\bibitem[{{Veras} {et~al.}(2017{\natexlab{a}}){Veras}, {Carter}, {Leinhardt},
  \& {G{\"a}nsicke}}]{Veras+2017_B}
{Veras}, D., {Carter}, P.~J., {Leinhardt}, Z.~M., \& {G{\"a}nsicke}, B.~T.
  2017{\natexlab{a}}, \mnras, 465, 1008

\bibitem[{{Veras} \& {Ford}(2010)}]{Veras}
{Veras}, D., \& {Ford}, E.~B. 2010, \apj, 715, 803

\bibitem[{{Veras} {et~al.}(2017{\natexlab{b}}){Veras}, {Georgakarakos},
  {Dobbs-Dixon}, \& {G{\"a}nsicke}}]{Veras+2017}
{Veras}, D., {Georgakarakos}, N., {Dobbs-Dixon}, I., \& {G{\"a}nsicke}, B.~T.
  2017{\natexlab{b}}, \mnras, 465, 2053

\bibitem[{{Veras} {et~al.}(2013){Veras}, {Mustill}, {Bonsor}, \&
  {Wyatt}}]{Veras+2013}
{Veras}, D., {Mustill}, A.~J., {Bonsor}, A., \& {Wyatt}, M.~C. 2013, \mnras,
  431, 1686

\bibitem[{{Veras} \& {Tout}(2012)}]{Veras+12}
{Veras}, D., \& {Tout}, C.~A. 2012, \mnras, 422, 1648

\bibitem[{{Veras} \& {Wolszczan}(2019)}]{Veras+2019}
{Veras}, D., \& {Wolszczan}, A. 2019, \mnras, 488, 153

\bibitem[{{Wang} {et~al.}(2019){Wang}, {Jones}, {Shporer}, {Fulton}, {Paredes},
  {Trifonov}, {Kossakowski}, {Eastman}, {Redfield}, {G{\"u}nther}, {Kreidberg},
  {Huang}, {Millholland}, {Seligman}, {Fischer}, {Brahm}, {Wang}, {Cruz},
  {Henry}, {James}, {Addison}, {Liang}, {Davis}, {Tronsgaard}, {Worku},
  {Brewer}, {K{\"u}rster}, {Zhang}, {Beichman}, {Bieryla}, {Brown},
  {Christiansen}, {Ciardi}, {Collins}, {Esquerdo}, {Howard}, {Isaacson},
  {Latham}, {Mazeh}, {Petigura}, {Quinn}, {Shahaf}, {Siverd}, {Rodler},
  {Reffert}, {Zakhozhay}, {Ricker}, {Vanderspek}, {Seager}, {Winn}, {Jenkins},
  {Boyd}, {F{\H{u}}r{\'e}sz}, {Henze}, {Levine}, {Morris}, {Paegert},
  {Stassun}, {Ting}, {Vezie}, \& {Laughlin}}]{Wang+2019}
{Wang}, S., {Jones}, M., {Shporer}, A., {et~al.} 2019, \aj, 157, 51

\bibitem[{{Xu} {et~al.}(2016){Xu}, {Jura}, {Dufour}, \& {Zuckerman}}]{Xu+2016}
{Xu}, S., {Jura}, M., {Dufour}, P., \& {Zuckerman}, B. 2016, \apjl, 816, L22

\bibitem[{{Xu} {et~al.}(2017){Xu}, {Zuckerman}, {Dufour}, {Young}, {Klein}, \&
  {Jura}}]{Xu+2017}
{Xu}, S., {Zuckerman}, B., {Dufour}, P., {et~al.} 2017, \apjl, 836, L7

\bibitem[{{Yan} {et~al.}(2018){Yan}, {Shi}, {Zhou}, {Chen}, {Li}, {Zhang},
  {Bi}, {Wu}, {Li}, {Guo}, {Liu}, {Gao}, {Zhang}, {Zhou}, {Li}, \&
  {Zhao}}]{Yan+2018}
{Yan}, H.-L., {Shi}, J.-R., {Zhou}, Y.-T., {et~al.} 2018, Nature Astronomy, 2,
  790

\bibitem[{{Zuckerman} {et~al.}(2011){Zuckerman}, {Koester}, {Dufour}, {Melis},
  {Klein}, \& {Jura}}]{Zuckerman+2011}
{Zuckerman}, B., {Koester}, D., {Dufour}, P., {et~al.} 2011, \apj, 739, 101

\bibitem[{{Zuckerman} {et~al.}(2003){Zuckerman}, {Koester}, {Reid}, \&
  {H{\"u}nsch}}]{Zuckerman+2003}
{Zuckerman}, B., {Koester}, D., {Reid}, I.~N., \& {H{\"u}nsch}, M. 2003, \apj,
  596, 477

\bibitem[{{Zuckerman} {et~al.}(2010){Zuckerman}, {Melis}, {Klein}, {Koester},
  \& {Jura}}]{Zuckerman+2010}
{Zuckerman}, B., {Melis}, C., {Klein}, B., {Koester}, D., \& {Jura}, M. 2010,
  \apj, 722, 725

\end{thebibliography}

\end{document}